# Pseudo-Riemannian metric: a new perspective on the quantum realm

Miaomiao Wei[1]†, Longjun Xiang[1]†, Fuming Xu[1*], Baigeng Wang[2,3], and Jian Wang[1,4*]

[1]College of Physics and Optoelectronic Engineering, Shenzhen University, Shenzhen 518060, China.

[2]National Laboratory of Solid State Microstructures and Department of Physics, Nanjing University, Nanjing 210093, China.

[3]Collaborative Innovation Center for Advanced Microstructures, Nanjing University, Nanjing 210098, China.

[4]Department of Physics, The University of Hong Kong, Pokfulam Road, Hong Kong, China.

† These authors contributed equally to this work.
*Correspondence to: xufuming@szu.edu.cn; jianwang@hku.hk.

**Abstract:**

As a fundamental concept in condensed matter physics, quantum geometry within the Riemannian metric elucidates various exotic phenomena, including the Hall effects driven by Berry curvature and quantum metric. In this work, we propose novel quantum geometries within a pseudo-Riemannian framework to explore unique characteristic of quantum matter. By defining distinct distances on pseudo-Riemannian manifolds and incorporating spin degree of freedom, we introduce the Pauli quantum geometric tensor. The imaginary part of this tensor corresponds to the Pauli Berry curvature, leading to the discovery a novel quantum phase: Pauli semimetal in $PT$-symmetric systems. This phase, characterized by the topological Pauli Chern number, manifests as a two-dimensional Pauli Chern insulator with helical edge states. These topological phases, uniquely revealed by the Pauli-Riemannian metric, go beyond the familiar Riemannian metric, where Berry curvature vanishes due to $PT$-symmetry. Pauli Chern number can classify helical topological insulator with or without time reversal symmetry. Pseudo-Riemannian metrics offer new insights into quantum materials and extend the scope of quantum geometry.

**Introduction**

Geometry is an indispensable tool for exploring the intricate and fascinating world of physics. Just as wearing different types of glasses can reveal various perspectives of the same scene, employing distinct geometries and metrics can also unveil diverse phenomena. For instance, Euclidean geometry is essential in Newtonian mechanics, whereas the Minkowski geometry within the pseudo-Riemannian metric is fundamental to the special relativity. Herein, we demonstrate that pseudo-Riemannian metrics can uncover unique phenomena in the quantum world, particularly compared with the Riemannian metric.

The exploration of spin as an internal degree of freedom dates back one century, and the analogy between charge and spin has been examined in various contexts such as charge current and spin current. Yet, this analogy can be further explored at a fundamental level to reveal novel physics. Quantum geometric tensor, which represents the distance in conventional Riemannian manifold and is associated with Berry curvature and quantum metric[1], has significantly advanced the field of quantum materials, particularly in the study of Hall effects[2,3], topological insulators (TIs) including both Chern insulators and $Z_2$ TIs[4-6], as well as topological semi-metals (SMs)[7,8]. Extending the concept of quantum geometry to a pseudo-Riemannian metric to further explore the analogy between charge and spin can yield intriguing insights. As a concrete example, we establish a new pseudo-Riemannian metric and obtain the following results.

Our work introduces a novel concept by defining the quantum "spin" distance between Bloch states in Hilbert space, establishing the Pauli quantum geometric tensor as a pseudo-Riemannian metric. The imaginary component of this tensor is identified as the Pauli Berry curvature, leading to the discovery of a new geometric phase, the Pauli Berry phase. This phase reveals a novel class of 2D helical TIs in $PT$-symmetric systems, termed the Pauli Chern insulator. Additionally, a new type of monopole charge is uncovered, characterizing systems where Berry curvature vanishes, thus defining the Pauli SM by endowing new topological characteristics to the Dirac SM. This framework naturally establishes a correspondence between the 3D Pauli SM and the 2D Pauli Chern insulator, while the Pauli Chern number, derived from the Pauli Berry curvature, serves as a universal topological invariant for helical TIs. These TIs exist across various symmetry classes, including $T$-symmetric, $T$-broken, and $PT$-symmetric systems. This finding suggests that the topological nature of helical TIs is deeply rooted in a pseudo-Riemannian manifold[9].

**Pauli quantum geometric tensor**

Following the treatment in Ref. 10, we calculate the distance between two points in Euclidean space which is generally defined through the inner product

$$D = \langle x|x \rangle = x_\mu x^\mu = \eta_{\mu\nu} x^\mu x^\nu, \quad (1)$$

where $\eta_{\mu\nu}$ is the metric tensor and the Einstein summation convention is implied. If the metric $\eta_{\mu\nu}$ is positive or semi-positive definite, it defines a Riemannian metric. Otherwise, the metric is pseudo-Riemannian (also called semi-Riemannian). One well-known example of the pseudo-Riemannian metric is the Minkowski metric, which is characterized by $\eta_{\mu\nu} = \bar{\eta}_\mu \delta_{\mu\nu}$ with $\bar{\eta}_\mu = (1, -1, -1, -1)$.

To derive quantum geometric tensor within the Riemannian metric, we consider the distance between two Bloch states $\psi_n(\boldsymbol{k})$ in the Hilbert space[10], where $x = \psi_n(\boldsymbol{k}+d\boldsymbol{k}) - \psi_n(\boldsymbol{k})$. Expanding $D$ up to the second order in $d\boldsymbol{k}$, we find $D = \langle \partial_\alpha \psi_n | \partial_\beta \psi_n \rangle dk_\alpha dk_\beta = \sum_m A^\alpha_{nm} A^\beta_{mn} dk_\alpha dk_\beta$, where $A^\alpha_{nm}$ is the Berry connection with $\alpha, \beta = x, y, z$. After removing the gauge-variant contribution[10], we obtain the quantum geometric tensor

$$g_n^{\alpha\beta} \equiv \sum_m r^\alpha_{nm} r^\beta_{nm}, \quad (2)$$

where $r^\alpha_{nm} \equiv A^\alpha_{nm}(1-\delta_{nm})$. Since the metric $\eta_{\mu\nu} = 1$ is used to derive Eq. (2), $g_n^{\alpha\beta}$ represents Riemannian description of the quantum world. In general, $g_n^{\alpha\beta}$ is a complex quantity, with its real component and imaginary component corresponding to the quantum metric and Berry curvature, respectively.

Alternatively, by defining a pseudo-Riemannian distance $D^\gamma \equiv \langle x | \sigma^\gamma | x \rangle$, where $\sigma^\gamma$ represents the Pauli matrix for spin, we derive the Pauli quantum geometric tensor

$$g_n^{\gamma,\alpha\beta} \equiv \sum_{ml} r^\alpha_{nm} \sigma^\gamma_{ml} r^\beta_{ln}, \quad (3)$$

as dictated by the Pauli-Riemannian metric $\eta = \sigma^\gamma$. Different from the pseudo−Riemannian Minkowski metric in the space-time coordinates, the Pauli-Riemannian metric focuses on the internal degree of freedom: spin. Further, by replacing the Pauli matrix $\sigma^\gamma_{ml}$ in Eq. (3) with the $2 \times 2$ identity matrix, we recover the Riemannian quantum geometric tensor Eq. (2) and hence the analogy between spin and charge particularly manifested in quantum geometry is exactly shown in Eqs. (2-3).

Similar to $g_n^{\alpha\beta}$, the imaginary part of Pauli quantum geometric tensor is the *Pauli Berry curvature*,

$$\bar{\Omega}_n^{\gamma,\alpha\beta} = -2 \sum_{ml} \text{Im}\left( r^\alpha_{nm} \sigma^\gamma_{ml} r^\beta_{ln} \right), \quad (4)$$

while its real part is the Pauli quantum metric $\bar{Q}_n^{\gamma,\alpha\beta} = \sum_{ml} \text{Re}\left( r^\alpha_{nm} \sigma^\gamma_{ml} r^\beta_{ln} \right)$.

Interestingly, there is a useful relation between Pauli Berry curvature and conventional Berry curvature[11]

$$\overline{\boldsymbol{\Omega}}_n^\gamma = \nabla_k \times \overline{\boldsymbol{\mathcal{A}}}_n^\gamma + \sigma_{nn}^\gamma \boldsymbol{\Omega}_n, \quad (5)$$

where $\overline{\mathcal{A}}_n^{\gamma,\alpha} = \sum_m \sigma_{nm}^\gamma r_{mn}^\alpha$, $\overline{\boldsymbol{\Omega}}_n^\gamma$ and $\boldsymbol{\Omega}_n$ are the vector form of Pauli and conventional Berry curvature, respectively. As shown in Table S1, Pauli Berry curvature exhibits completely different symmetry transformations compared to conventional Berry curvature. For example, in systems with $PT$ symmetry, Berry curvature $\boldsymbol{\Omega}_n$ vanishes, whereas Pauli Berry curvature $\overline{\boldsymbol{\Omega}}_n^\gamma$ may exist and reveal unique topological features.

Consequently, we have $\overline{\boldsymbol{\Omega}}_n^\gamma = \nabla_k \times \overline{\boldsymbol{\mathcal{A}}}_n^\gamma$ in $PT$-symmetric systems ($\boldsymbol{\Omega}_n = 0$). This form highlights an analogy between Berry physics and Pauli Berry physics, suggesting potential discovery of a novel monopole that could lead to a new geometric phase. It further suggests the possible existence of topological phases (Pauli topological SMs and Pauli TIs). For this purpose, we define the Pauli Chern number in 2D as

$$C^\gamma = \sum_n \int_k \hat{\boldsymbol{z}} \cdot \overline{\boldsymbol{\Omega}}_n^\gamma, \quad (6)$$

where $\hat{\boldsymbol{z}}$ is the normal direction of the 2D system and $\int_k$ stands for the integral for the whole Brillouin zone. Pauli Chern number allows us to characterize the topological properties of quantum systems where conventional Berry curvature fails to provide insights, thereby expanding our understanding of topological phases. In the rest of this paper, unless otherwise specified, we will focus on $T$-broken but $PT$-symmetric systems.

The Pauli Berry curvature can be generalized by replacing $\sigma^\gamma$ in Eq. (4) with a general non-positive definite metric matrix $\mathcal{M}$, which shares the same dimension of the Hamiltonian. For a four-band model, where the Hamiltonian is spanned by Pauli matrices $\sigma^\gamma$ and $\tau^\alpha$, the metric matrix $\mathcal{M}$ can also be expanded on the basis $\tau^\alpha \sigma^\gamma$, where $\alpha, \gamma = 0,1,2,3$, resulting in 15 components of (generalized) Pauli Berry curvature (Note that $\tau^0 \sigma^0$ gives the Riemannian Berry curvature). From now on, we will use $\sigma^\alpha$ to denote Pauli matrices for real spin and $\tau^\alpha$ to represent the Pauli matrices for pseudo-spin. Throughout this paper, $\tau^\alpha \sigma^\gamma$ denotes $\tau^\alpha \otimes \sigma^\gamma$.

Since the energy spectrum of $PT$-symmetric system is doubly degenerate, we need to discuss the non-Abelian Pauli Berry curvature in a way similar to the Riemannian Berry curvature. As discussed in detail in the Supplementary Infromation[11], non-Abelian Riemannian Berry curvature is a $2 \times 2$ matrix for doubly degenerate systems and ***only its trace has physical meaning***[1]. It is straightforward to show that non-Abelian Berry curvature can be obtained by excluding the degenerate states in the definition of Berry curvature. Similarly, we can express the non-Abelian Pauli Berry curvature as

$$\overline{\Omega}_{n_\varsigma}^{\gamma,\alpha\beta} = -\sum_{m_\kappa l_\xi \neq n_\pm} 2\mathrm{Im}\left(r_{n_\varsigma m_\kappa}^\alpha \sigma_{m_\kappa l_\xi}^\gamma r_{l_\xi n_\varsigma}^\beta\right), \quad (7)$$

where $\kappa, \xi, \varsigma = \pm$ denote the doubly degenerate states. With Eq. (7), we can calculate Pauli Chern number for doubly degenerate systems. As shown in Fig.1(c), the topological features of Dirac SMs are somewhat ill-defined within the Riemannian metric, particularly due to the vanishing of Berry curvature and the absence of monopole charge. In contrast, as demonstrated below, Pauli Berry curvature defined in the Pauli-Riemannian manifold can possess a monopole charge and hence can reveal unique topological properties of Dirac SMs.

**Pauli semimetal**

Topological SMs, including Weyl SMs and Dirac SMs, represent a class of exotic gapless topological quantum materials[7]. In particular, Weyl SMs, characterized by two-fold band crossing points, can only manifest in 3D crystalline solids without either $P$ or $T$ symmetry. In contrast, Dirac SMs, defined by four-fold band degeneracy points, can only exist in materials with $PT$ symmetry. In addition to the gapless Dirac cones at the Fermi level in the bulk, both Weyl and Dirac SMs exhibit surface Fermi arc states that terminate at the projections of the gapless Weyl or Dirac nodes on that surface, as illustrated in Fig. 1(a-c). Typically, these topological surface states can be understood through the quantum geometric language defined on a Riemannian manifold, such as the Berry curvature.

In the following, we introduce a novel topological SM, termed Pauli SM, which is described by the Pauli-Riemannian metric. This Pauli SM can manifest in $PT$-symmetric systems where the existence of Weyl SMs is prohibited. The minimal model for a $PT$- symmetric antiferromagnetic system has the following four-band Hamiltonian[12], $H = \sum_{i=0}^{5} d_i(\mathbf{k})\Gamma^i$, with $\Gamma^0 = \tau_0\sigma_0$, $\Gamma^1 = \tau_x\sigma_0$, $\Gamma^2 = \tau_z\sigma_0$, $\Gamma^3 = \tau_y\sigma_x$, $\Gamma^4 = \tau_y\sigma_y$, $\Gamma^5 = \tau_y\sigma_z$. For this minimal model, we define

$$H = \mathbf{d} \cdot \mathbf{\gamma}, \quad (8)$$

where $\mathbf{d} = (d_x, d_y, d_z)$ and $\mathbf{\gamma} = (\gamma_x, \gamma_y, \gamma_z) = (\Gamma^i, \Gamma^j, \Gamma^k)$. There are totally ten possible Pauli SMs (for $k > j > i$), which are listed in Table S2 of the Supplementary Information[11]. Assuming $\mathbf{d} = \mathbf{k}$ in Eq. (8), for all the members in Table S2, we find that $\bar{\mathbf{\Omega}}_n = -\mathbf{k}/(2k^3)$ and $\nabla_k \cdot \bar{\mathbf{\Omega}}_n = 2\pi\kappa_n\delta(\mathbf{k})$, where $\kappa_n = \pm 2$ is the monopole charge and $n$ labels the energy level which are doubly degenerate. In general, the $PT$-symmetric Hamiltonian can be divided into spin-conserved and spin-nonconserved groups, where the typical members are exemplified in Table I, along with their topological characteristics. Note that different metric matrices $\mathcal{M}$ have been used for different members in Table S2.

We emphasize that the physical entities of Dirac SMs and Pauli SMs are the same; the difference lies in the physical interpretation. Thus, there is only one Fermi arc for Eq. (8). Both Dirac SMs and Pauli SMs refer to the same Hamiltonian in Eq. (8), but are analyzed with different metrics----Riemannian for Dirac SMs and Pauli-Riemannian for Pauli SMs. It in turn ensures the stability of the Pauli SM based on Eq. (8), where the stability of Dirac points have been discussed in Ref. 12. In addition, Pauli SM in $PT$-symmetric systems, attributes new topological origin for the experimentally realized Dirac SMs Na$_3$Bi[13] and Cd$_3$As$_2$[14].

The chirality and its associated monopole charge of Pauli SM can be introduced through the Hamiltonian defined in Eq. (8) with $\boldsymbol{\gamma} = (\Gamma^1, \Gamma^2, \Gamma^5)$ and $\boldsymbol{d} = \left(k_x, k_y, M(k_w^2 - k^2)\right)$, where $k^2 = k_x^2 + k_y^2 + k_z^2$. Here $k_w$ labels the band crossing points and $M$ is a tunable system parameter[15]. The monopole charge of Pauli SMs at $(0, 0, \pm k_w)$ is calculated as $\pm 2$, which doubles the monopole charge found in Weyl SMs. As a result, two edge states appear when the 3D Pauli SM is projected onto a 2D $PT$-symmetric TI.

**Pauli Chern insulator, a new 2D helical TI**

Considering the Hamiltonian in Eq. (8) with $\boldsymbol{\gamma} = (\Gamma^1, \Gamma^2, \Gamma^5)$. As we discussed previously, when $\boldsymbol{d} = \boldsymbol{k}$, it is a Pauli SM. When it is gapped into a 2D system, from the tight-binding Hamiltonian $\boldsymbol{d} = \left(sink_x, sink_y, M - cosk_x - cosk_x\right)$, we find $C^x = 2$ for $M \in (-2, 0)$ and $C^x = -2$ for $M \in (0, 2)$, as shown in Fig. 2(b), which indicates the topological nature of the new TI. Fig. 2(a) gives the band structure of this new TI when confined in the $k_y$ direction, where two degenerate band crossings are observed in the gap. In Fig. 2(c), we depict the two-probe transport coefficient using the Landauer-Büttiker formula where the quantized conductance persists in the presence of weak Anderson disorder demonstrating the robustness of the new TI. In Fig. 2(d), the quantum spin Hall conductance is plotted against disorder strength $W$ indicating the helical nature of the new TI. The details of the calculation are shown in the Supplementary Information[11]. Since this system is obtained when a Pauli SM is gapped, we term it as Pauli Chern insulator due to the breaking of $T$ symmetry.

**Fermi arc of Pauli semimetal**

Next, we discuss the physical origin of the surface Fermi arc states of Pauli SMs. In a $T$-broken Weyl SM, each Weyl node carries a monopole charge of $\pm 1$ as indicated by the red (positive) and blue (negative) ball in Fig. 1(a), respectively. When a plane cutting through the line connecting the Weyl nodes (assuming $k_y$ direction) it receives a net quantized flux, allowing the plane to be identified as a 2D Chern insulator with chiral edge states, see Fig. 1(b). By scanning $k_y$ between the Weyl nodes, one finds that their projections on the $k_x$-$k_y$ plane are connected by the surface Fermi arc[16], see Fig. 1(a). One sees that the existence of monopole is crucial for the appearance of surface Fermi arc states in Weyl SMs.

Similarly, in the minimal model of Pauli SM with two nodes at $(0, 0, \pm k_w)$, the monopole charge is $\pm 2$, arising from the non-Abelian Pauli Berry curvature[11]. Like in Weyl SM, a double Fermi arc is observed on the surface of Pauli SM, as shown in Fig.

1(c). By inserting a plane between these nodes, a Pauli Chern insulator with helical edge states emerges, as illustrated in Fig. 1(e).

**Characterizing helical TIs with Pauli Chern number**

Besides $PT$-symmetric systems, we show that the Pauli Chern number can serve as a new topological invariant, equivalent to the $Z_2$ invariant for $Z_2$ TIs ($T$-symmetric), and can also be applied to characterize magnetic TIs ($T$-broken)[11].

The first model is the Kane-Mele model[17,18] for quantum spin Hall (QSH) insulator (with or without $P$ symmetry) and the second model is the Bernevig-Hughes-Zhang (BHZ) model for HgTe/CdTe quantum wells[19]. Both models have $T$ symmetry and are classified as $Z_2$ TIs. The third model is a magnetic TI (MTI)[15,20] with $T$-broken symmetry, where no topological invariant is available for its QSH state. The robustness of QSH phase for $T$-broken systems has been confirmed in several works[21,22]. We find that Pauli Berry curvature with metric $\mathcal{M} = \tau_z \sigma_0$ can be used to characterize the topological invariant for this QSH state. As summarized in Table II, for all helical states with $T, PT$, and $T$-broken symmetries, we have $C^\gamma = 2$.

**Discussion**

The pseudo-Riemannian metrics can be extended to include various internal and external degrees of freedom. For instance, by substituting spin with parity (particle-hole), a parity-Riemannian metric can be introduced to describe quantum matter in superconducting states. Be- yond the global geometry explored in this work, the pseudo-Riemannian quantum geometry can also be applied to investigate the response properties of quantum matter in metallic states. This can involve examining local pseudo-Riemannian quantum geometric features such as Pauli Berry curvature and Pauli Berry curvature dipole, further expanding the scope of quantum geometric investigations.

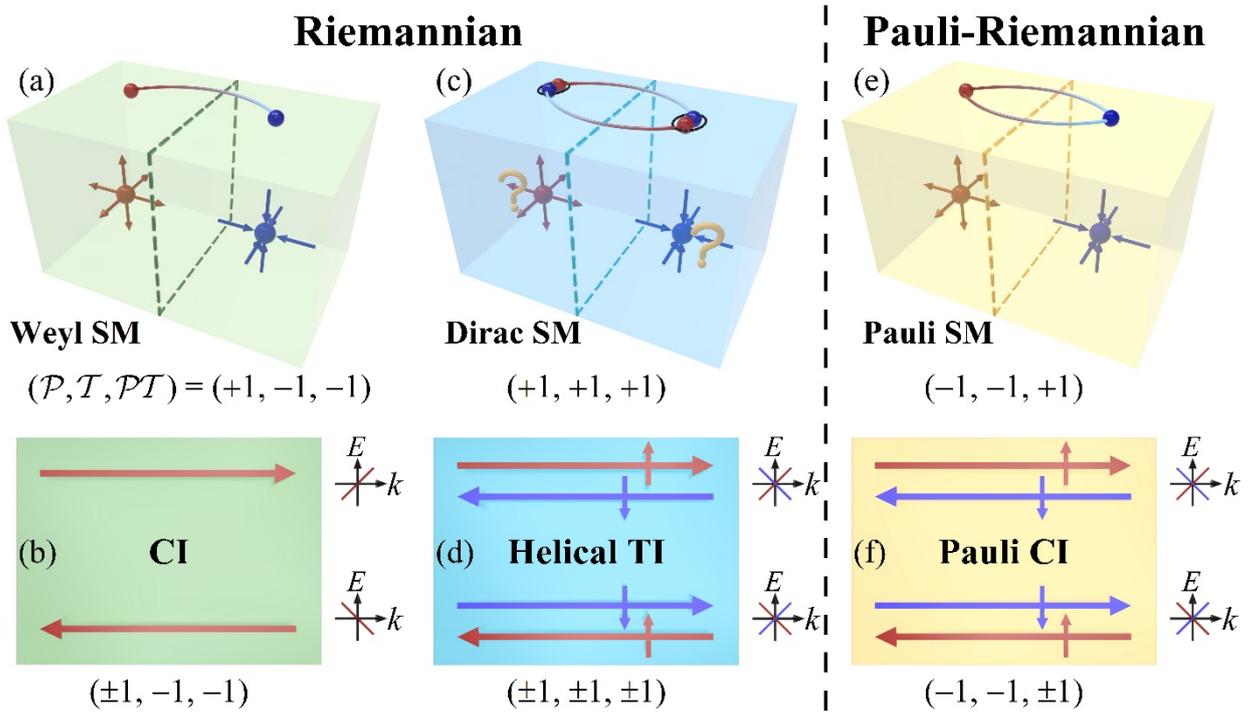

**Figure 1:** The correspondence between three-dimensional (3D) topological semimetals and 2D TIs. Here $(P, T, PT) = (\pm 1, \pm 1, \pm 1)$ denotes inversion ($P$), time-reversal ($T$), and $PT$ symmetries, where $+1$ ($-1$) represents the presence (absence) of the symmetry. (a) and (b): 3D Weyl SMs without $T$ and 2D Chern insulator (CI). (c) and (d): 3D Dirac SMs with $PT$ symmetry and 2D helical TIs. Due to the PT symmetry, the net monopole charge of Dirac point is zero and hence the correspondence between 3D Dirac SM and 2D helical TI is illusive especially when spin is not conserved. (e) and (f): 2D Pauli Chern insulator can be acquired from the two Dirac points of 3D Pauli SMs without $P$ and $T$ but with $PT$. The surface Fermi arc states (indicated by the red and grey lines) of 3D topological SMs feature similar physical origin with the edge states (denoted by the red and blue arrows) of 2D TIs.

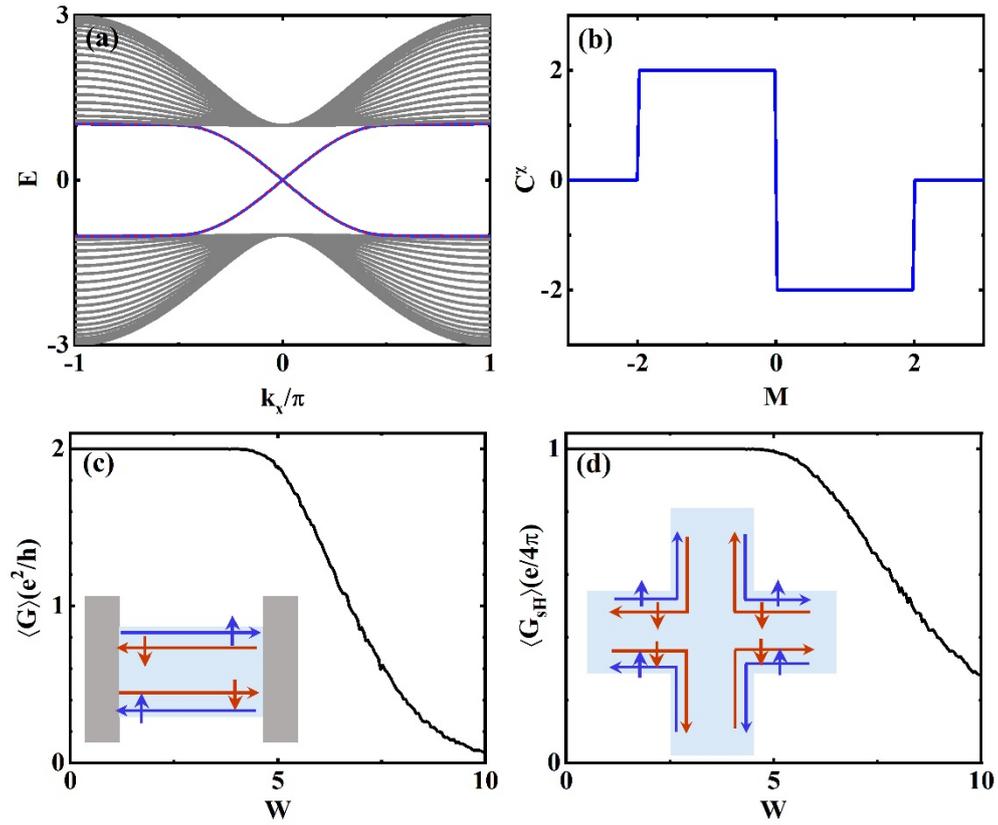

**Figure 2: Pauli Chern insulator.** (a) The band structure with a four-fold degenerate band crossing. (b) Pauli Chern number as a function of the system parameter M. (c) Disorder-average conductance for a two-probe setup, where a robust plateau against disorder is observed. (d) Disorder-average quantum spin-Hall conductance for a four-probe setup.

**Table 1:** In typical Pauli SMs with the $PT$-symmetric Hamiltonian $H = \boldsymbol{d} \cdot \boldsymbol{\gamma}$, a vector $\boldsymbol{d} = (k_x, k_y, k_z)$ gives rise to a monopole charge $\kappa_n = 2$ for all Pauli SMs (*11*). When gapped into a 2D system, these systems transform into helical TI with a nonzero Pauli Chern number $|C^\gamma| = 2$ when $M \in (-2,2)$. From a Riemannian perspective, Dirac SMs, which consist of a pair of degenerate Weyl nodes labeled $+$ and $-$, can be characterized by the 2D Chern numbers for the doubly degenerate levels when gapped, denoted as $(C_+, C_-)$.

|  | Spin-conserved | Spin-nonconserved |
|---|---|---|
| $(\gamma_x, \gamma_y, \gamma_z)$ | $(\Gamma^1, \Gamma^2, \Gamma^5)$ | $(\Gamma^3, \Gamma^4, \Gamma^5)$ |
| $\mathcal{M}$ | $\tau_0 \sigma_z$ | $\tau_y \sigma_0$ |
| $(C_+, C_-)$ | $(1, -1)$ | $(0, 0)$ |
| $C^\gamma$ | 2 | 2 |

**Table 2:** Pauli Chern number for different models of helical TIs. NA means not applicable.

| Model | Kane-Mele[17,18] | BHZ[19] | MTI[15] | Pauli CI [Eq. (8)] |
|---|---|---|---|---|
| Phase | $Z_2$ helical | $Z_2$ helical | QSH | QSH |
| Symmetry | $T$ | $T$ | $T$-broken | $PT$ ($T$-broken) |
| Riemannian topological invariant | $Z_2 = 1$ | $Z_2 = 1$ | NA | NA |
| Pauli Chern number | $C^z = 2$ | $C^z = 2$ | $C^y = 2$ | $C^x = 2$ |

**Acknowledgments:** This work was supported by the National Natural Science Foundation of China (Grants No.12034014 and No.12174262).

**Author contributions:** J.W. conceived the project. J.W. and L.J.X. developed the theory. M.W. and F.X. performed the transport calculation. L.J.X., F.X., B.W., and J.W. wrote the paper. All authors contributed to the discussion of the results.

**Competing interests:** The authors declare no competing interests.

**Data and materials availability:** Data in the main text and supplementary information, as well as the codes for theoretical calculations are available from the corresponding authors upon reasonable request.